\documentclass[aip,apl,reprint,amssymb,amsmath,amsfonts,superscriptaddress]{revtex4-1}
\usepackage[utf8]{inputenc}
\usepackage[T1]{fontenc}
\usepackage[american]{babel}
\usepackage{graphicx}
\usepackage{float}
\usepackage[hidelinks]{hyperref}

\begin{document}

\preprint{1}
\title{Lingering dynamics in microvascular blood flow}

\author{A. Kihm}
\affiliation{Department of Experimental Physics, Saarland University, 66123 Saarbruecken, Germany}

\author{S. Quint}
\affiliation{Department of Experimental Physics, Saarland University, 66123 Saarbruecken, Germany}
\affiliation{Cysmic GmbH, 81379 München}

\author{M. W. Laschke}
\affiliation{Institute for Clinical and Experimental Surgery, Saarland University, 66421 Homburg, Germany}

\author{M. D. Menger}
\affiliation{Institute for Clinical and Experimental Surgery, Saarland University, 66421 Homburg, Germany}

\author{T. John}
\affiliation{Department of Experimental Physics, Saarland University, 66123 Saarbruecken, Germany}

\author{L. Kaestner}
\affiliation{Department of Experimental Physics, Saarland University, 66123 Saarbruecken, Germany}
\affiliation{Theoretical Medicine and Biosciences, Saarland University, 66421 Homburg, Germany}

\author{C. Wagner}
\email[]{c.wagner@uni-saarland.de}
\affiliation{Department of Experimental Physics, Saarland University, 66123 Saarbruecken, Germany}
\affiliation{Physics and Materials Science Research Unit, University of Luxembourg, Luxembourg}


\date{\today}

\begin{abstract}
The microvascular networks in the body of vertebrates consist of the smallest vessels such as arterioles, capillaries, and venules. The flow of RBCs through these networks ensures the gas exchange in as well as the transport of nutrients to the tissues. Any alterations in this blood flow may have severe implications on the health state. Since the vessels in these networks obey dimensions similar to the diameter of RBCs, dynamic effects on the cellular scale play a key role. 
The steady progression in the numerical modeling of RBCs, even in complex networks, has led to novel findings in the field of hemodynamics, especially concerning the impact and the dynamics of lingering events, when a cell meets a branch of the network. However, these results are yet to be matched by a detailed analysis of the lingering experiments in vivo. To quantify this lingering effect in in vivo experiments, this study analyzes branching vessels in the microvasculature of Syrian golden hamsters via intravital microscopy and the use of an implanted dorsal skinfold chamber.
It also presents a detailed analysis of these lingering effects of cells at the apex of bifurcating vessels, affecting the temporal distribution of cell-free areas of blood flow in the branches, even causing a partial blockage in severe cases.
\end{abstract}

\maketitle

\section*{Introduction}
The steady transport of nutrients to the tissues of the body, as well as the delivery of oxygen, is crucial for the health state in animals and humans. However, in unicellular living beings, this task can be achieved by pure diffusion. Active transport is necessary to ensure this exchange in vertebrates. This process is realized by the synergistic action of the heart, blood, and vasculature \cite{kaestner_lars_calcium_2013}. A key setting is the pressure difference between the aorta and the vena cava \cite{popel_microcirculation_2005}. In between, the microvasculature is present, consisting of the smallest vessels such as the arterioles, capillaries, and venules. The diameters of these vessels in the microcirculation are in the range of the size of individual RBCs, which highlights the importance of the deformation ability of RBCs. Due to the overall length of the microvasculature and the present cross-sections of vessels, it is responsible for the largest resistance and hence the largest dissipation of energy in the blood flow \cite{poiseuille_recherches_1832,pries_resistance_1994,gould_capillary_2017}. Since the microcirculation is embedded in the tissues and ensures the perfusion of these and the organs, any alteration in blood flow may have severe consequences on the actual health state. Indeed, several 
pathological states are linked to the disturbance in the microcirculation, such as Alzheimer's disease \cite{gutierrez-jimenez_disturbances_2018,de_la_torre_evidence_2000}. Angiogenesis and angioadaptation are hereby prone to fulfil the metabolic needs of the respective organs and tissues \cite{pries_making_2014,pries_design_1995}. The observed architectures range from tree-like networks to mesh-like networks, where the subsequent diameters of vessels at bifurcations satisfy, in general, Murray's law, although deviations are well-known \cite{murray_physiological_1926,sherman_connecting_1981}. 
Compared to macrovascular blood flow, effects arising from the particulate nature of blood are more pronounced in the microcirculation. Considering the volume fraction of RBCs in whole blood suspensions, the so-called hematocrit, a temporal heterogeneity, is given in the capillary vessels contrasting a temporal homogeneity ubiquitous in large vessels such as arteries (cf. Movie S2). Similarly, the vessel diameter in which the blood flows has been found to have an impact on the local hematocrit \cite{fahraeus_suspension_1929,pries_a_r_blood_1990}. This so-called F\aa hr\ae us effect can be explained by the lateral migration of RBCs to the vessel centerline, leaving a cell-depleted layer close to the vessel walls. As a result, RBCs obey a higher speed on average in the Poiseuille profile than the bulk speed of the plasma. In the seminal work by F\aa hr\ae us and Lindqvist \cite{fahraeus_viscosity_1931}, the former effect could be identified as one of the major causes of the dependence on the vessel diameter of apparent viscosity in blood solutions. Apart from these observations, phase separation in the microvasculature becomes apparent, leading to a nonhomogeneous distribution of RBCs within the vessels. This phase separation is partially caused by the F\aa hr\ae us effect, especially in conjunction with bifurcations, albeit it is not the only origin. The entirety of the described effects highlighting the significant impact of the bi-phasic composition of blood also show the necessity to model blood in this way.\\
Recent advances in the modeling of the microvasculature in mammals have led to findings of the dynamics of RBCs, as well as other cellular components in silico. The complexity of the networks in these studies ranges from one symmetrical bifurcation \cite{ye_motion_2019,bacher_antimargination_2018} to the mimicking of an anatomically accurate in vivo network consisting of an interconnected mesh-like structure with various branches, confluences, and bifurcations \cite{pozrikidis_numerical_2009,bagchi_mesoscale_2007,balogh_computational_2017,balogh_direct_2017}. In ref. \cite{balogh_direct_2017}, such complex networks have been investigated on a scale that  both allows for dense suspensions of RBCs as well as yielding discrete datapoints of every single RBC, including their individual shape. As a result, deformation and dynamic characteristics of RBCs in the vicinity of bifurcations have been studied, leading to the observation of the so-called lingering events. This phenomenon of RBCs resting at the apex of bifurcations is well known in physiology; however, to our knowledge, no systematic studies have been carried out to address this phenomenon. Further, the results found in ref. \cite{balogh_direct_2017} are still unmatched by any means in vivo. Thus, this study aims to elucidate the fundamental dynamics of these lingering events in vivo. By means of intravital fluorescence microscopy, it can extract position data out of flowing RBCs in the microcirculatory system of hamsters. Additionally, we have developed algorithms that allow to separate the effect of lingering on the flow of subsequent RBCs. Specifically, we use this approach to show the impact of lingering on the void duration, i.e., cell-free areas in the bloodstream.

\section*{Materials and Methods}
\subsection*{Permissions}
All the conducted experiments were approved by the local government animal protection committee (permission number: 25/2018) and were performed in accordance with the German legislation on the protection of animals and the NIH Guidelines for the Care and Use of Laboratory Animals.
\subsection*{Animal preparation}
We briefly describe the necessary steps in the preparation protocol (for a more detailed description, see \cite{laschke_dorsal_2011}). Syrian golden hamsters (\emph{Mesocricetus auratus}) with a bodyweight of 60--80\,g are equipped with a dorsal skinfold chamber, consisting of two symmetrical titanium frames with a total weight of approx. 4\,g. For this purpose, the animals are anesthetized and from their depilated and disinfected back, one layer of skin and subcutis with the panniculus carnosus muscle, as well as the two layers of the retractor muscle, are completely removed within the area of the observation window of the chamber. A total of five hamsters were prepared to neglect any interindividual effects and to observe different phenomena in various geometries.
\subsection*{Experimental setup}
For intravital microscopic analyses of the microcirculation, the hamster is anesthetized by an intraperitoneal injection of 100\,mg/kg ketamine and 10\,mg/kg xylazine, and 0.1\,ml of the blood plasma marker 5\,\% fluorescein isothiocyanate (FITC)-labeled dextran (150 kDa, Sigma-Aldrich, Taufkirchen, Germany) is injected into the retrobulbar venous plexus for contrast enhancement. Subsequently, the animal is fixed on a stage, allowing for horizontal positioning of the desired field of view under the objective of an upright microscope (Zeiss AG, Oberkochen, Germany), as previously described for mice \cite{hertz_evolution_2019}. Due to the geometry of the observation window (circular with a diameter of 10\,mm) and the attached snap ring of the chamber, the use of liquid immersion objectives is provided. To maximize the observation area, one ideally uses narrow objectives since the objective may collide with the frame by examining an area close to the boundaries of the observation window.\\
In this study, two different objectives are used: a water immersion objective for investigations of capillary blood flow with a magnification of $63\times$ and an air objective with a magnification of $50\times$ (both Zeiss). Its high magnification allows for tracking and analyzing individual RBCs (e.g., the cell shape evolution while passing through confluences or bifurcations). On the contrary, an air objective with increased working distance is used to record blood flow in vessel geometries of bigger dimensions (enlarged field of view) or embedded in deeper tissue layers. 
The recorded image series consist of up to 6,000 images, leading to a time coverage of $4-20$\,s depending on the actual frame rate of the camera (ORCA-Flash4.0 V3, Hamamatsu Photonics K.K., Hamamatsu, Japan).
\subsection*{Image processing}
Due to the inherent properties of a dynamical system, the amount of fluorescent dye in the field of view is highly time-dependent. This fact results in a flickering motion in the original footage. Histogram matching has been applied for uniform white balance throughout the images and to suppress these flickering events. A Gaussian blur was then applied to despeckle the images before applying a binary mask to disregard the background and enhance the contrast of the image series. An example of this mask is depicted in Fig.\,(\ref{fig:geometry_signal}), where the original in vivo geometry can be found along with the corresponding mask, created by averaging all images and tracing the resulting mean image. Particle tracking has been carried out via a custom-tailored \textsc{Matlab}\textsuperscript{\textregistered}(MathWorks, Massachusetts, USA) script (see also \cite{guckenberger_numericalexperimental_2018,bacher_antimargination_2018}). By this technique, we were able to extract the position data of the moving cells. We stress that due to abundant breathing movements, not all images of a series can be analyzed but rather split into subseries of images where no displacements of the vessels are visible. However, alterations in tissue thickness may lead to deficiencies in image quality. The same holds for the fact that vessels are winding in a three-dimensional topology, and thus, the focal plane will only capture a certain part of the geometry. Therefore, a complete automatized analysis is complicated and possible in only some peculiar cases. For most analyses, manual adjustments and evaluations must be carried out.
Since the plasma in the hamsters is stained with a fluorescent dye, the apparent RBCs obey lower brightness values than cell-free (plasma rich) areas. From the corresponding temporal brightness distributions, one can therefore define voids and the passing RBCs. For uniform characterization of void durations, we binarized the signal with respect to the mean value, i.e., all the signals above the mean will be considered as void. In each branch, the mean passage time of RBCs are determined and denoted by $\tau_\text{RBC}$. For a better comparison within a geometry, we seek to achieve a normalized void duration. Therefore, we have divided the calculated void durations by $\tau_\text{RBC}$ in the respective branch to achieve a normalization by the flow rate. These normalized void durations are then sorted in ascending order to obtain the empirical cumulative distribution function. We postulate this empirical cumulative distribution function to be represented by a log-normal distribution function $\text{cdf}\left(\frac{\tau_n}{\tau_\text{RBC}}\right)$, 
\begin{align}
	\text{cdf}\left(\frac{\tau_n}{\tau_\text{RBC}}\right)=\frac{1}{2}\left[1+\text{erf}\left(\frac{\log\left(\frac{\tau_n}{\tau_\text{RBC}}\right)-\hat{\mu}}{\sqrt 2\hat{\sigma}}\right)\right]\,,\label{eq:cdf}
\end{align}
with the error function $\text{erf}(\cdot)$, and parameters $\hat{\mu},\hat{\sigma}\in\mathbb{R},\hat{\sigma}>0$.
These paramaters are estimated based on our dataset using the maximum-likelihood approach, yielding 
\begin{align}
	\hat{\mu} &= \frac{1}{N}\sum\limits_{n=1}^{N}\log\left(\frac{\tau_n}{\tau_\text{RBC}}\right)\,,\nonumber\\\hat{\sigma}^2 &= \frac{1}{N\!-\!1}\sum\limits_{n=1}^N \left(\log\left(\frac{\tau_n}{\tau_\text{RBC}}\right)-\hat{\mu}\right)^2\,.\label{eq:est_parameters}
\end{align}
Probability density distributions of the normalized void durations are given by differentiation of Eq.\,[\ref{eq:cdf}]. The postulation of the log-normal distribution of normalized void durations has been verified a posteriori by a Kolmogorov-Smirnov test. For the geometry in Fig.\,\ref{fig:geometry_signal}, the corresponding cumulative densities of void durations are depicted in Fig.\,\ref{fig:CDF}.

\section*{Results}
\begin{figure*}[!htb]
	\centering
	\includegraphics[width=\textwidth]{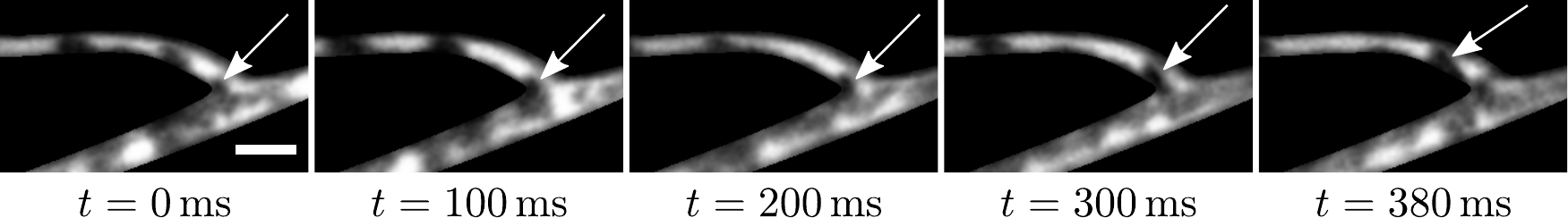}
	\caption{Time series of a lingering RBC at an arterial bifurcation for a time interval of $t=380$\,\text{ms}. The plasma was fluorescently labeled, and therefore, RBCs appear as dark spots. At $t=0$\,\text{ms}, an RBC is touching the apex of the bifurcation, marked by the arrow. The cell starts to deform and linger around this apex, leading to a partial blockage with decreased flow rate, as can be seen in the upper daughter vessel for all subsequent images. Finally, the cell is detached from the apex at $t=380$\,\text{ms}. The scale bar is 10\,$\mu$m in width. Additional data is provided in Fig.\,S1.2.}
	\label{fig:lingering_timeseries}
\end{figure*}

\begin{figure*}[t]
	\centering
	\includegraphics[width=\linewidth]{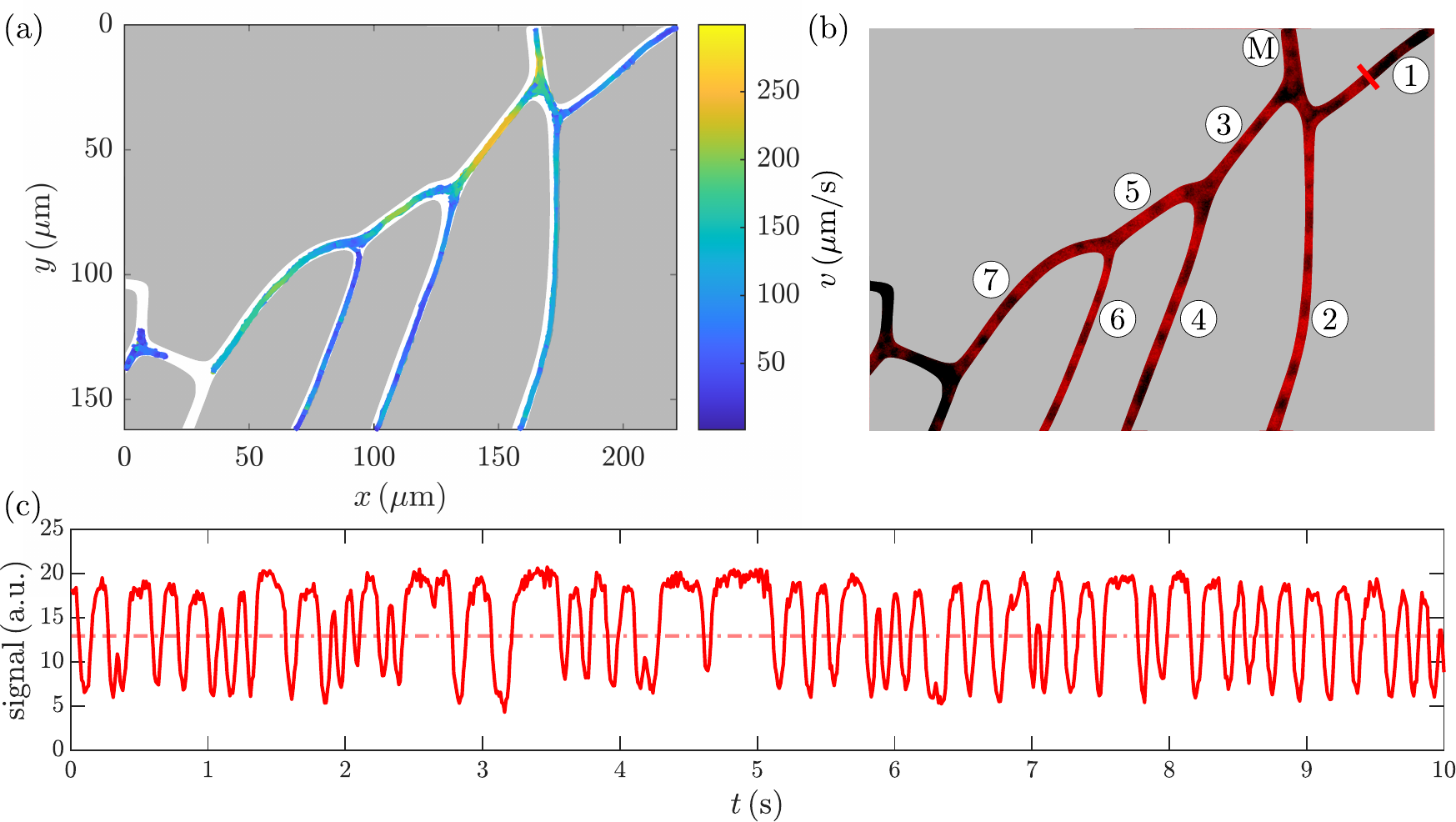}
	\caption{(a) Results of particle tracking of RBCs in the given geometry. The flow is coming from top (mother vessel ``M'' in Figure (b)) and exits in all other branches. The colorbar corresponds to the tracked velocities and depicted is the superposition of 500 tracks. (b) Snapshot of the geometry with flowing RBCs (red). To enhance the contrast and visibility, false color images are shown. The daughter branches are labeled in ascending order from rightmost to leftmost and will be referred to in the main text. (c) Distribution of voids within a branch (1) of a bifurcation. The graph corresponds to the measured integrated intensity along a perpendicular line segment with respect to the centerline of this branch (red line segment in branch (1) in (b)). Values above the mean value (red dash-dotted line) can be regarded as voids, i.e., an absence of cells, whereas values below the mean value correspond to passing cells. On average, voids have a duration of approx. 100\,ms; however, due to partial blockage caused by lingering RBCs, void formation can exceed multiple times the average duration, as can be seen at $t\approx 4.7\,$s, where a void with a duration of $350\,$ms is formed.}\label{fig:geometry_signal}
\end{figure*}

We analyze bifurcating vessels in the microvascular system of hamsters (arterioles, capillaries, and venules) with varying diameter and bifurcating angles (details are given in the Materials and Methods section). A typical scenario of a lingering event is shown in Fig.\,\ref{fig:lingering_timeseries}, where the temporal evolution of a lingering RBC in an arteriolar bifurcation has been recorded. 
We analyzed a variety of different geometries and hamster models (see Supplementary Information). Most of the analyzed geometries exhibit one apex with two branching vessels; however, we have presented here the interesting data of the more complex geometry with four apices and a total of seven branching vessels (see Fig.\,\ref{fig:geometry_signal}).
Based on the geometry depicted in Fig.\,\ref{fig:geometry_signal}, we calculate the integrated brightness signal along a line perpendicular to the respective centerline of the vessel. 

To take the lingering into account, we further applied a particle-tracking algorithm yielding trajectories of individual RBCs. Out of these tracking data, we can extract detailed knowledge of the RBC velocities. Since a lingering event is defined by an RBC resting at the apex of a bifurcation, we analyzed the velocity data in a small region around the apex of the respective bifurcation. If, in this region, the speed of passing RBCs obeys a severe drop, we call this a lingering event. The lingering duration is quantized as a time interval when $v_\text{RBC} < 30\,\mu\text{m}/\text{s}$. This value is significantly lower than that of typical cell speeds in the microvasculature, which are in the range of $v_\text{RBC}=100\,\mu\text{m}/\text{s}$. We did not set this lingering speed to zero because the detected center of mass may shift slightly in consecutive images according to fluctuations. We want to stress that the cause of this adaptation is exclusively due to the experimental nature and does not contradict a lingering event, as defined in \cite{balogh_direct_2017}.

In general, the combined application of both the analysis of brightness signals and particle tracking is needed since it cannot be guaranteed that the trajectories cover the whole distance the single RBCs are travelling due to limited image resolution. On the other hand, the sole evaluation of brightness signals along the vessel centerline will not be sufficient to detect lingering events due to the complex dynamics of RBCs at the apex. Thus, we used a combination of both techniques in the sense that we analyzed void durations by evaluating brightness signals and applying the particle tracking data as a filter to separate the influence of lingering on the distribution of voids in each branch of the given microvascular network.
Due to experimental restrictions, we have to deal with not all visible parts of a given vessel being situated in the focal plane since they are exploiting a three-dimensional topology. To overcome this drawback, we obtained the previously described cumulative brightness signal at a vessel segment that is in focus and thus corrected for the spatio-temporal shift of lingering RBCs at the bifurcation apex and the influence on the flow field thereof at a position further downstream. This shift is computed by the average flow speed in the vessel segment. As a result, the impact of distinct lingering scenarios can be associated with the formation of voids at a given position in the daughter branches.
\begin{figure}[!htb]
	\centering
	\includegraphics[width=.49\textwidth]{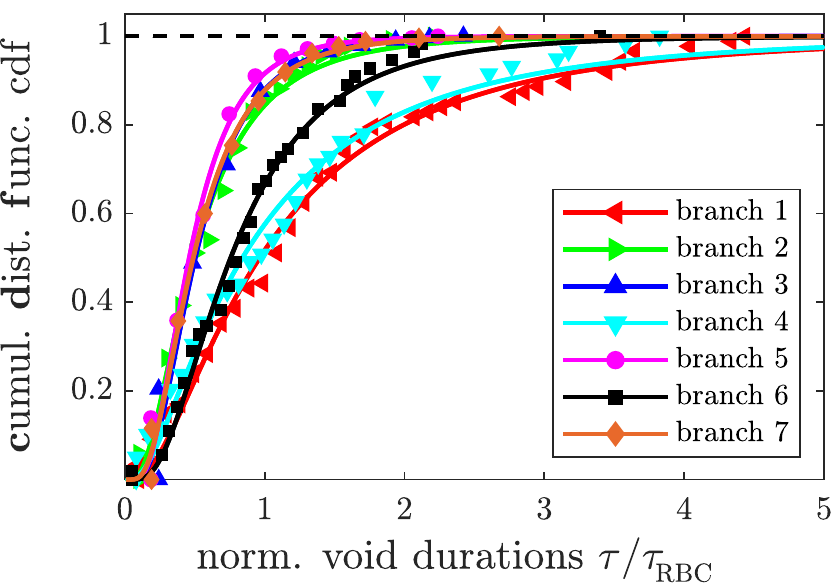}
	\caption{Cumulative distribution functions of void durations $\tau$ for all branches of the geometry in Fig.\,\ref{fig:geometry_signal}. The temporal length of the voids is hereby scaled for each branch by the average time of a RBC to pass, $\tau_\text{RBC}$. The data points correspond to measured void durations, whereas the solid line corresponds to the respective log-normal distributions with estimated parameters $\hat{\mu}$ and $\hat{\sigma}$, as in Eq.\,(\ref{eq:cdf}).}
	\label{fig:CDF}
\end{figure}
Using the maximum-likelihood approach, we estimated the parameters of our empirical distribution, cf. Materials and Methods section. In Fig.\,\ref{fig:CDF}, the good agreement between the dataset and the estimated cumulative distribution function is shown. The corresponding probability density functions of voids for the geometry in Fig.\,\ref{fig:geometry_signal} are given in Figs.\,\ref{fig:PDF_nonlingering} and \ref{fig:PDF_lingering}, resp.
\begin{figure}[!htb]
	\centering
	\includegraphics[width=.49\textwidth]{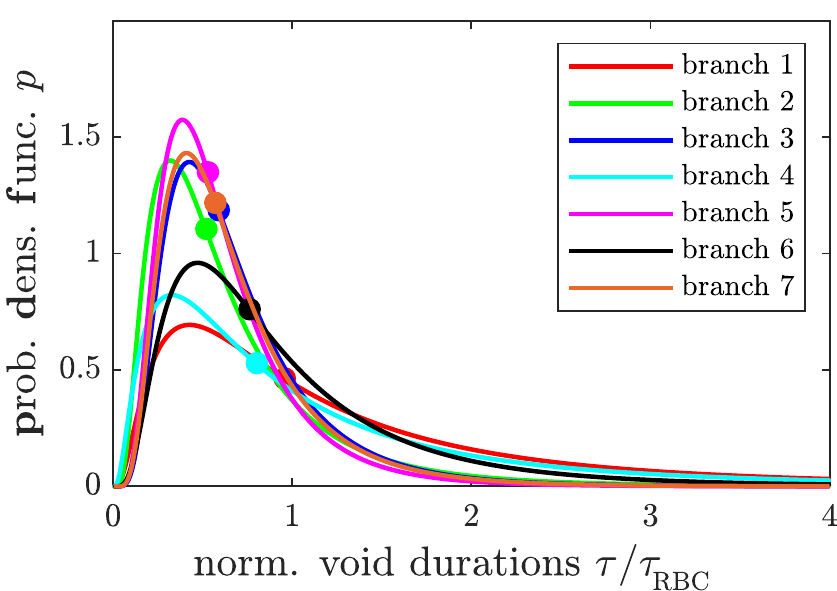}
	\caption{Probability density functions of void durations for all branches as in Fig.\,\ref{fig:CDF} in the case of non-lingering events. The temporal length of the voids is hereby scaled for each branch by the average time of a RBC to pass, $\tau_\text{RBC}$. Median values obtained from estimated parameters in Eq.\,[\ref{eq:est_parameters}] are indicated by filled circles in the respective color code.}
	\label{fig:PDF_nonlingering}
\end{figure}
\begin{figure}[!htb]
	\centering
	\includegraphics[width=.49\textwidth]{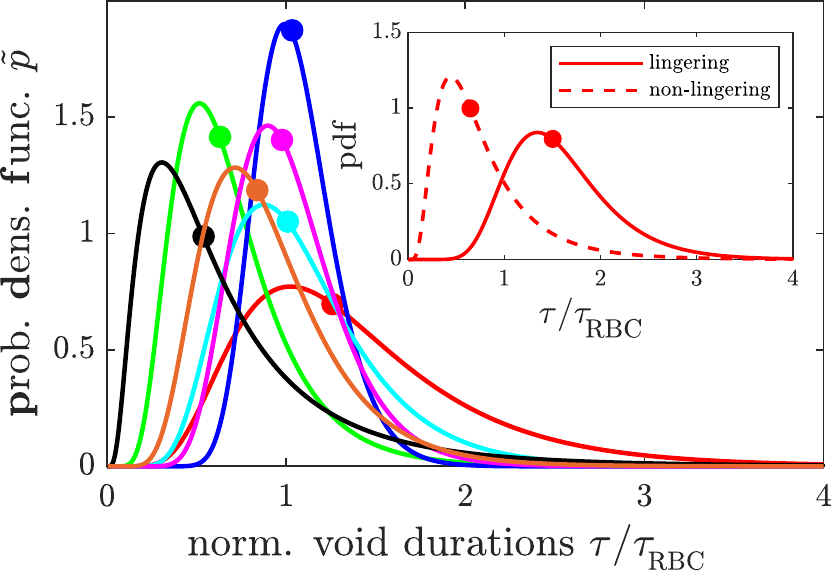}
	\caption{Probability density functions of scaled void durations for all branches if only lingering events are taken into account. We define a lingering event to occur if the speed of an RBC is lower than $v_\text{RBC}\leq 30\mu$m/s in the vicinity of a bifurcation apex. The legend is identical to the one in Fig.\,\ref{fig:CDF}. The inset graph shows both the probability densities in the case of lingering and non-lingering, respectively, for the geometry in Fig.\,\ref{fig:lingering_timeseries} to represent extreme cases. Additional information about this geometry can also be found in Fig.\,S1.2. Filled circles in matching colors denote median values of normalized void durations, obtained from estimated parameters in Eq.\,[\ref{eq:est_parameters}].}
	\label{fig:PDF_lingering}
\end{figure}
In the first case, only void durations associated to non-lingering events were taken into account, whereas void durations exclusively associated with lingering events were taken into account in the latter case. By comparing the graphs for each branch in both the figures, the influence of lingering on the void durations is obvious. In Fig.\,\ref{fig:PDF_nonlingering}, the median values of the empirical probability distrubution for all graphs are narrowly distributed. Contrasting this state, for voids associated with lingering events (Fig.\,\ref{fig:PDF_lingering}), we find a shift of medians toward higher void durations. A severe case of this observation can also be seen in the inset graph of Fig.\,\ref{fig:PDF_lingering}, where the median void duration was more than double in the lingering case with respect to non-lingering events.

\begin{figure}[!htb]
	\centering
	\includegraphics[width=.49\textwidth]{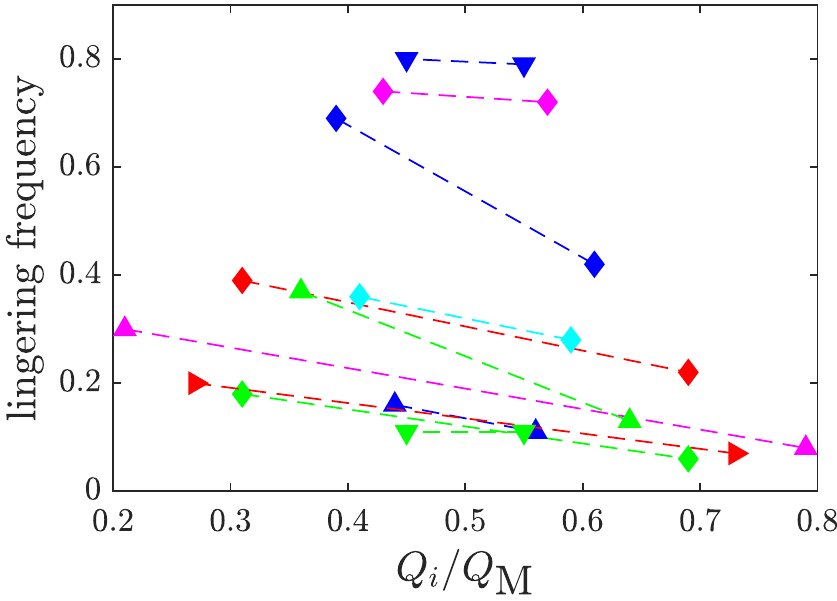}
	\caption{Lingering frequencies of the detected voids in relation to the normalized mean flowrate in a distinct vessel. The lingering frequency is hereby defined as the fraction of the void count associated with a lingering event and the total void count in the vessel. Further, we define the normalized mean flowrate as a fraction of the flowrate in a daughter vessel $Q_i$ and the mother vessel $Q_\textnormal{M}$. Identical color codes belong to pairs of vessels branching from the same apex; the dashed lines connect the data points of vessels.}
	\label{fig:lingering_frequency}
\end{figure}

\begin{figure}[!htb]
	\centering
	\includegraphics[width=.49\textwidth]{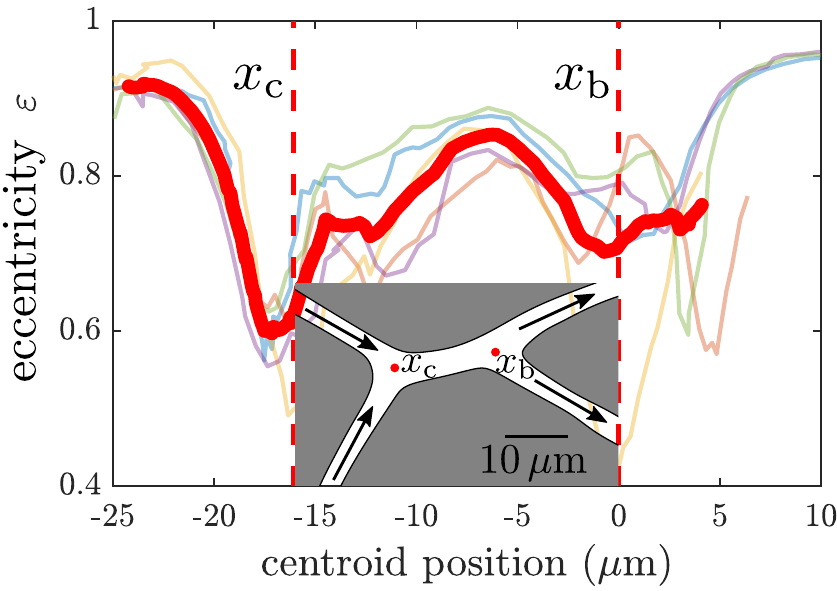}
	\caption{Eccentricity $\varepsilon$ of RBCs as a function of the centroid position within the geometry shown as the inset. The eccentricity is hereby calculated as the ratio of the distance between the two foci and the length of its major axis of an ellipse with identical second moments for each individual RBC for all consecutive images. The thick red solid line represents the average of all individual graphs (thin lines). For the analysis, only single RBCs are considered, whereas trains of flowing RBCs are neglected. The offset of the centroid position is chosen in a way that the bifurcation apex $x_\text{b}$ is at position zero.}
	\label{fig:eccentricity}
\end{figure}

In addition to the probability densities of void durations, we can also define a so-called lingering frequency as the fraction of voids not associated with lingering events and the total number of occurring voids in a branch of the network. Fig.\, \ref{fig:lingering_frequency} shows the calculated lingering frequencies of all the analyzed vessels in relation to the normalized mean flowrate in the respective vessel. The normalization factor is given by the mean flowrate of the mother or feeding vessel, which is equal to the sum of the flowrates of all draining vessels due to the incompressibility of the fluid. Even though the size of RBCs is comparable to the apparent vessel diameters, their speed may serve as a good approximation of the mean speed of the surrounding fluid (plug flow); hence, we find 
\begin{align}
	Q=\Delta V/\Delta t=A\,v_\textnormal{fluid}\simeq A\,l_\textnormal{RBC}/\tau_\textnormal{RBC}\,,
\end{align}
with the time-averaged flowrate $Q$, the volume element $\Delta V$, the cross-sectional area $A$, mean speed of the fluid $v_\text{fluid}$, the length of the major axis of the circumscribing ellipse of RBCs $l_\textnormal{RBC}$ and the average cell passage time $\tau_\textnormal{RBC}$, as introduced in previous paragraphs. Among all the investigated pairs of bifuracting vessels, we find the lingering frequency to be higher in the one with lower flowrates with respect to its counterpart with a higher flowrate.
Apart from lingering at bifurcations, RBCs may also deform in the vicinity of branches, i.e., bifurcations or confluences. While lingering implies the strong interaction of the vessel walls, the sole presence of junctions may induce shape changes for approaching or distancing RBCs. The main difference is the reduction of the speed, which is significant in the case of lingering, but adapted to the flow rates in the respective branch in the latter case, although slight deviations may occur. To analyze the spatio-temporal evolution of this deformation, we calculated the circumscribing ellipse for each individual RBC for all consecutive images, yielding both the centroid position as well as the eccentricity of the cell, given as the ratio of the distance between the two foci and the length of its major axis. Fig.\,\ref{fig:eccentricity} shows the measured eccentricity values for the flowing RBCs in the confluence-bifurcation geometry, both individually as well as the average curve. The corresponding geometry exhibiting both a bifurcation and a confluence is shown as an inset of Fig.\,\ref{fig:eccentricity}. From the average curve, one can clearly see the transition of cell shapes RBCs undergo while flowing. At the position of the confluence apex $x_\text{c}$, the mean eccentricity exhibits the global minimum, implying the roundest obtained shape. Similarly, at the position of the bifurcation apex $x_\text{b}$, the mean eccentricity exhibits a local minimum, indicating a transformation from an elongated to a more spherical shape when approaching the apex and again elongating when entering one daughter branch.

\section*{Discussion}
Since we analyzed the void formation downstream in a bifurcating vessel geometry, the apparent increase in median void durations originates from two possible scenarios. One contribution is given by the redistribution of consecutive RBCs into the adjacent daughter vessel. The second contribution is given by a change in void speed due to an altered flow rate in the vessel that has an impact on the temporal void duration. In the latter case, the spatial distance between the consecutive RBCs would sustain, outmatching the observed state. Indeed, by considering the standard deviation of the speeds of passing RBCs in a vessel, variations of the flow rates are negligible. Thus, temporal void durations and spatial void lengths are highly correlated, and this implies a breakup of clusters of RBCs approaching a bifurcation apex. The term cluster hereby implies the state of RBCs moving in a chain where the intercellular distance is in the order of the cellular size, where hydrodynamic interactions are abundant \cite{claveria_clusters_2016,brust_plasma_2014}. However, we emphasize that the increase of the median void durations in the case of lingering RBCs does not hold for all the analyzed geometries (cf. Supplementary Materials).

Further, we also notice a suppression of very short void durations, as can be seen from the comparison of Figs.\,\ref{fig:PDF_nonlingering} and \ref{fig:PDF_lingering}. To quantify this statement, we calculated the probabilities for void durations being less than $0.5\,\tau_\text{RBC}$. In the case of non-lingering, integration of the corresponding probability densities yields for the probabilities 
$P_i\left(\tau_\text{void}<0.5\,\tau_\text{RBC}\right) = \{ 0.23, 0.47, 0.38,\allowbreak 0.31, 0.45, 0.27, 0.40\},$ $i\in\{1,\dots,7\}$, where $i$ denotes the branch identifier according to Fig.\,\ref{fig:geometry_signal}. Similarly, we obtain $\tilde{P}_i\left(\tau_\text{void}<0.5\,\tau_\text{RBC}\right) = \{ 0.02, 0.30, 0.00, 0.03, 0.01, 0.46,\allowbreak 0.10\},$ $i\in\{1,\dots,7\}$ with the probabilities $\tilde{P}_i$ in the case of lingering. If one compares these values for each branch, it is obvious that void durations less than or equal to $0.5\,\tau_\text{RBC}$ are suppressed drastically in all but for $i=6$.

In some of the graphs showing the probability densities of the void durations, rather long-tailed distributions are present. We stress that these tails arise inherently due to the heterogeneous distribution of RBCs in the microvascular networks, leading to cell-depleted sequences in branches and thus long void durations in absence of lingering cells.

One crucial question is the dependence of the lingering frequency on the flow properties. Fig.\,\ref{fig:lingering_frequency} shows a decreased frequency for the branch with the higher flowrate with respect to the adjacent vessel. In the prevalent low Reynolds number regime, RBCs follow merely the streamlines of the surrounding plasma, and thus, one finds fewer cells in the vessel transporting less volume. However, the interaction of cellular compounds with the endothelial walls of the vessels is complex \cite{pries_microvascular_2005}, and therefore, it is highly non-intuitive to observe this circumstance. It is even more remarkable given the broad distribution of opening angles in all the analyzed geometries. For normalized flowrates close to $0.5\,Q_\text{M}$, we obtained very similar lingering frequencies for both the connected vessels. Nevertheless, the overall magnitude of the lingering frequency seems to be unaffected by this observation and rather depends on the cutting angle between the daughter vessels of the geometry in a way that small angles exhibit higher lingering frequencies than large ones in the majority of cases. Other flow parameters such as absolute flow rates or curvature of the bifurcation apex may also influence the lingering frequency.   

So far, we have focused on the impacts of lingering RBCs on the microvascular blood flow in vivo. However, the physical prerequisites to obey lingering have not been discussed yet. RBCs obey an inner network of spectrin fibers, as they are responsible for their biconcave shape at rest. Due to the flexibility of this spectrin network, RBCs can pass through constrictions much smaller than their size at rest \cite{mohandas_mechanical_1994,heinrich_elastic_2001}. Yet, not only constrictions alter the shape of RBCs, but also the complex structure of the vascular network itself, exhibiting merging and bifurcating vessels. The shape of RBCs undergoes a characteristic deformation when approaching the apex of a bifurcation or a confluence, respectively (cf. Fig.\,\ref{fig:eccentricity}). Recently, this behavior was reproduced in silico for a varying number of passing cells \cite{ye_motion_2019}. Whereas this alteration of the shape is due to increasing or decreasing confinements depending on the geometry, it is responsible for the observed lingering behavior. Particles such as hard spheres obey a less severe coupling with the fluid, and we assume the deformation and the strong fluid-cell interaction of RBCs is the major cause of lingering \cite{happel_low_1965}.

\section*{Conclusion}
We used cutting edge intravital microscopy in conjunction with a combined sophisticated signal processing algorithm and particle tracking to obtain detailed information of flowing RBCs in living hamster models. Based on this data we define and detect so called lingering events, i.e. RBCs resting at a bifurcation apex of branching vessels. We show, that these lingering events particularly cause a redistribution of subsequent RBCs in the adjacent daughter vessels and lead to a break-up of trains of RBCs. We further analyze the ratio of lingering cells and all traversing RBCs, the so called lingering frequency, which is found to be higher in the branching vessel with the higher flow rate compared to the adjacent vessel.
All the presented results of our study show a good qualitative agreement with in silico results in ref.\,\cite{balogh_direct_2017} although, in contrast to the well-defined boundary conditions in silico, the major experimental drawback is the limited insight into the whole model system. These limitations are given by a limited field of view and the sheer complexity of the living hamster model and all its parameters. Nevertheless, we can assess the impact of lingering RBCs on the flow behavior of subsequent cells in vivo. We can provide evidence to show that these lingering events cause a breakup of trains of RBCs as well as redistribution in the branching vessels. Even though these effects seem to be rather fine-grained, the impact on the whole organism may be severe, given the importance of blood flow to the health state. 



\section*{Author Contributions}

M.W.L., M.D.M., L.K., and C.W. designed the research; A.K., M.W.L., and S.Q. performed the research; A.K., and T.J. analyzed the data; A.K. and C.W. wrote the paper; and M.W.L, S.Q., M.D.M., L.K., and T.J. provided feedback and insights.

\section*{Acknowledgments}

AK, TJ, LK, and CW gratefully acknowledge support from the research unit DFG FOR 2688 - Wa1336/12 of the German Research Foundation. MWL and MDM received support from the research unit DFG FOR 2688 - LA2682/9-1 of the German Research Foundation. This work was supported by the European Union’s Horizon 2020 research and innovation programme under the Marie Sk\l{}odowska-Curie grant agreement No. 860436 – EVIDENCE (SQ, LK, CW). CW, TJ, SQ, and AK kindly acknowledge the support and funding of the “Deutsch-Französische-Hochschule” (DFH) DFDK CDFA-01-14 “Living fluids”.

\section*{Supplementary material}
The final manuscript including all supplementary data is accessible via \href{https://doi.org/10.1016/j.bpj.2020.12.012}{doi:10.1016/j.bpj.2020.12.012}.

\section*{Acknowledgments}

AK, TJ, LK, and CW gratefully acknowledge support from the research unit DFG FOR 2688 - Wa1336/12 of the German Research Foundation. MWL and MDM received support from the research unit DFG FOR 2688 - LA2682/9-1 of the German Research Foundation. This work was supported by the European Union’s Horizon 2020 research and innovation programme under the Marie Sk\l{}odowska-Curie grant agreement No. 860436 – EVIDENCE (SQ, LK, CW). CW, TJ, SQ, and AK kindly acknowledge the support and funding of the “Deutsch-Französische-Hochschule” (DFH) DFDK CDFA-01-14 “Living fluids”.

\bibliography{bibliography_lingering} 

\end{document}